\documentclass{ws-procs9x6}

\begin{document}

\title{Study of nuclear giant resonances using a Fermi-liquid method}

\author{Bao-Xi Sun}

\address{Institute of Theoretical Physics, College of Applied
Sciences, \\ Beijing University of Technology, Beijing 100124, China \\
E-mail: sunbx@bjut.edu.cn}

\begin{abstract}
The nuclear giant resonances are studied by using a Fermi-liquid method,
and the nuclear collective excitation
energies of different values of $l$ are obtained, which are fitted
with the centroid energies of the giant resonances of spherical
nuclei, respectively. In addition, the relation between the isovector giant resonance and the corresponding
isoscalar giant resonance is discussed.
\end{abstract}


\bodymatter

\section*{}

In nuclear physics, the Landau parameters are derived
microscopically from the ground state energy in the relativistic
mean-field approximation, and Landau Fermi liquid theory is used to
describe the thermodynamics properties of the nuclear systems, such
as the compressibility, the symmetric energy and the hydrodynamics
sound velocities\cite{matsui81}. This theory is also studied with an
effective chiral Lagrangian to obtain the properties of the nuclear
ground state and the link between an effective QCD theory and the
nuclear many body theory\cite{Friman96,Song01}. Moreover, the low
momentum nucleon-nucleon potential is applied to calculate the
effective interaction between quasi-particles near the Fermi
surface, and then the static properties of the nuclear matter are
extracted\cite{Kuo07}.

Landau Fermi liquid theory is an important method to describe the
low energy collective excitation properties of many-body systems,
and it can be used to study the problem on giant resonances of nuclei,
which is still an hot topic in nuclear physics~\cite{Greiner}.
With several typical methods, such as the random
phase approximation with Skyrme interactions\cite{Zhang}, the
relativistic random phase approximation\cite{Cao,Ma,Ring09}, the
centroid energies and strength distributions of the giant resonances
of nuclei are calculated and compared with the experimental
data.

In this work, I will try to calculate the collective excitation
energy of the nuclear system within a Fermi-liquid model\cite{Wen},
which is generalized to the three-dimension situation with the spin of nucleons included\cite{Sun1003}.
In the calculation, a Lagrangian of Walecka model is used as a microscopic input\cite{Walecka}.
The calculation results will be compared with the
experimental data of the nuclear giant resonance of some spherical nuclei.
Furthermore, I hope to check whether the contribution of Dirac sea of
nucleons is essential in the study of collective excitations of nuclear
many-body systems.

The liquid equation of motion of the quasi-nucleon in the momentum space can be written as
\begin{eqnarray}
\label{eq:Schrodinger}    i\frac{\partial}{\partial t}
u_\alpha(l,\vec{q},t) &=&Hu_\alpha(l,\vec{q},t), \nonumber \\
\end{eqnarray}
where the Hamiltonian
\begin{eqnarray}
\label{eq:hamilton} H_{l,l^\prime} &=& q \left(a_{l} \delta_{l+1,l^\prime} ~+~a_{l-1,}
\delta_{l-1,l^\prime} \right) \nonumber \\&&
\left(v^\ast_F-\frac{k^2_F}{(2\pi)^3}f_F(l,l)\right)^{1/2}\left(v^\ast_F-\frac{k^2_F}{(2\pi)^3}f_F(l^{\prime},l^{\prime
})\right)^{1/2}
\end{eqnarray}
is hermite with
\begin{equation}
a_{l}~=~\sqrt{\frac{(l+1)^2}{(2l+1)(2l+3)}} \nonumber
\end{equation}
and $v^\ast_F$ the Fermi velocity of nucleons in the relativistic mean-field approximation.
The eigenvalues of $H$ would give us
the frequencies of the collective excitation modes of the nuclear
matter.
Moreover, it can be seen from Eq.~(\ref{eq:hamilton}) that the exchange interaction between nucleons $f_F(l,l)$ causes
the collective excitation of the nuclear matter directly.

The parameters in
Ref.~\cite{HS} are used in the calculation, i.e., $g_\sigma=10.47$,
$g_\omega=13.80$, $m_\sigma=520MeV$, $m_\omega=783MeV$ and
$M_N=939MeV$. Since the nucleon near the Fermi surface would be more
possible to be excited, we set the value of  nucleon momentum
$|\vec{q}|=k_F=1.36fm^{-1}$ in the calculation. When $f_F(l,l)=0$,
the Hamiltonian $H$ has a continuous spectrum and it generates the
particle-hole continuum of the nuclear matter in the relativistic
mean-field approximation. However, if the value of $f_F(l,l)$ is
large enough, in addition to the continuum eigenvalues, the spectrum
of $H$ has isolated positive and negative eigenvalues. 
Actually, the
mode with a positive eigenvalue corresponds to the creation of a
nuclear collective excitation mode, while the the mode with a
negative eigenvalue corresponds to the annihilation of a nuclear
collective excitation mode.

The nuclear isoscalar giant resonances actually correspond to the
nuclear collective excitations with different values of $l$.
However, the nuclear isovector giant resonances correspond to the
nuclear collective excitation states that the collective excitation
of protons with the energy $E_S(l)$ is creating , while the
collective excitation of neutrons with the energy
$E_S(l)$ is annihilating, and vice versa. Hence, the energy of the nuclear isovector
giant resonance is about twice of the corresponding isoscalar giant
resonance in the nuclear matter, i.e.,
\begin{equation}
\label{twotime} E_V(l)=E_S(l)-(-E_S(l))=2E_S(l).
\end{equation}
The experimental data on the nuclear giant resonances in
Ref.~\cite{Cao74,Cao83,Cao84,Cao80,Cao81,Cao82} demonstrate that the
relation between the energy of nuclear isovector giant resonance and
that of nuclear isoscalar giant resonance in Eq.~(\ref{twotime}) is
correct approximately except for the case $l=1$, which will be
discussed in detail specially.  In follows, the
giant resonance energies with different values of $l$ in the
nucleus will be calculated within the Fermi-liquid model.

The nuclear giant monopole mode, $l=0$,  is also called as ${\sl breathing~mode}$ of the nucleus.
Supposed the proton and neutron densities can be calculated
approximately:
 \begin{equation}
\rho_p = \rho_0 \frac{Z}{A},~~~~\rho_n = \rho_0 \frac{N}{A},
\end{equation}
the calculated energies for isoscalar and isovector giant
monopole resonances of nuclei $^{208}Pb$, $^{144}Sm$, $^{116}Sn$,
$^{90}Zr$, $^{40}Ca$ and their corresponding experimental values are
listed in Table~\ref{tab:monopole}. Since the Fermi momentum of
protons $k_F(p)$ is different from that of neutrons, the collective
excitation energies of protons and neutrons, $E_0(p)$ and $E_0(n)$,
are different from each other.
When the effective nucleon mass takes a large value, i.e., $M^\ast_N=0.742M_N$, the calculation results of
the proton excitation energy for heavy nuclei, such as $^{208}Pb$,
$^{144}Sm$ and $^{116}Sn$, are fitted with the corresponding
experimental centroid energies of the nuclear isoscalar monopole
resonance $E^S_{exp}$, respectively.
However, for those light nuclei, $^{90}Zr$ and $^{40}Ca$, we must reduce the effective nucleon mass to $M^\ast_N=0.717M_N$, and then the reasonable calculation results fitted with the experimental data are obtained.
Moreover, the sum of the
excitation energies of protons and neutrons $E_0(p)+E_0(n)$ should
correspond to the nuclear isovector giant monopole energy.
In Table~\ref{tab:monopole}, We can find that calculated values of isovector giant monopole energies for $^{208}Pb$,  $^{90}Zr$ and $^{40}Ca$ are
in the range of the experimental errors.

\begin{table}
\tbl{The Fermi momenta and the $l=0$ collective excitation
energies of protons and neutrons for different nuclei with different
effective nucleon masses. 
The corresponding experimental values for the nuclear isoscalar
 giant monopole resonances $E^S_{exp}$ are taken from Ref.~\cite{Cao74}, and the
experimental values for the nuclear isovector giant monopole
resonances $E^V_{exp}$ from Refs.~\cite{Cao80,Cao81,Cao82}.}
{\begin{tabular}{@{}cccccc@{}}\toprule
$l=0$               & $^{208}Pb$  & $^{144}Sm$           & $^{116}Sn$        & $^{90}Zr$      & $^{40}Ca$ \\
\colrule
$M^\ast_N/M_N$      &   0.742  &  0.742  & 0.742  & 0.717 & 0.717 \\
$k_F(p)~(fm^{-1})$  &  $1.26$  &  $1.29$ & $1.29$ & $1.31$ &$1.36$ \\
 $k_F(n)~(fm^{-1})$ &  $1.45$  & $1.42$ &  $1.42$ & $1.41$ &$1.36$ \\
 $E_0(p)~(MeV)$     &  $16.28$ & $15.26$ & $15.26$ &$17.57$ & $15.58$ \\
  $E_0(n)~(MeV)$    &  $7.05$  &  $9.00$ &  $9.00$ &$13.13$ &  $15.58$ \\
$E_0(p)+E_0(n)~(MeV)$ &$23.33$ &$24.26$ &  $24.26$ &$30.7$ & $31.16$ \\
  $E^S_{exp}~(MeV)$ &  $14.17\pm0.28$ & $15.39\pm0.28$ & $16.07\pm0.12$ & $17.89\pm0.20$ & $-$      \\
  $E^V_{exp}~(MeV)$ &  $26.0\pm3.0$ & $-$ &   $-$ & $28.5\pm2.6$ &    $31.1\pm2.2$ \\
\botrule
\end{tabular}
}
\label{tab:monopole}
\end{table}

The dipole deformation of the nucleus is really a shift of the
center of mass. Thus the isovector giant dipole resonance of
the nucleus actually corresponds to the creation of the $l=1$
collective excitation of one kind of nucleons, i.e., protons or neutrons. However, the isoscalar giant
dipole resonance in $^{208}Pb$ with a centroid energy at
$E=22.5MeV$, which is discovered by studying the $(\alpha,\alpha^\prime)$ cross sections at
forward angles\cite{Cao77}, should be a compression mode, and
corresponds to a creation of the $l=1$ collective excitation of
protons or neutrons and an annihilation of the $l=1$ collective
excitation of the other sort of nucleons, i.e., neutrons or protons simultaneously. The calculation
results and the corresponding experimental centroid energies are
listed in Table~\ref{tab:dipole}. Similarly, we must adjust the effective
nucleon mass to obtain the collective excitation energies fitted with the experimental data.
It is apparent that in order to obtain a more correct
excitation energy, the effective nucleon mass must take a larger
value for heavy nuclei, but a smaller value for light nuclei.
For the nucleus $^{208}Pb$, the summation of the collective excitation
energies of protons and neutrons is equal to the experimental energy of
the isoscalar giant dipole resonance, approximately.

\begin{table}
\tbl{The Fermi momenta and the $l=1$ collective excitation
energies of protons and neutrons for different nuclei with different
effective nucleon masses. 
The corresponding experimental values for the nuclear isoscalar
 giant dipole resonances $E^S_{exp}$ are taken from Ref.~\cite{Cao77}, and the
experimental value for the nuclear isovector giant dipole resonance $E^V_{exp}$
from Ref.~\cite{Cao83}.}
{\begin{tabular}{@{}cccccc@{}}\toprule
$l=1$               & $^{208}Pb$  &  $^{90}Zr$      & $^{40}Ca$ \\
\colrule
$M^\ast_N/M_N$      &   0.755  &   0.742 & 0.70 \\
$k_F(p)~(fm^{-1})$  &  $1.26$  &   $1.31$ &$1.36$ \\
 $k_F(n)~(fm^{-1})$ &  $1.45$  &  $1.41$ &$1.36$ \\
 $E_1(p)~(MeV)$     &  $15.53$ & $15.56$ & $19.58$ \\
  $E_1(n)~(MeV)$    &  $6.57$  & $11.37$ &  $19.58$ \\
$E_1(p)+E_1(n)~(MeV)$ &$22.1$   &$26.93$ & $39.16$ \\
  $E^S_{exp}~(MeV)$ &  $22.5$       &  $-$          & $-$      \\
  $E^V_{exp}~(MeV)$ &  $13.5\pm0.2$ &  $16.5\pm0.2$ &    $19.8\pm0.5$ \\
\botrule
\end{tabular}
}
\label{tab:dipole}
\end{table}

The calculation results and the corresponding experimental centroid
energies of the giant quadrupole resonances of different nuclei are
listed in Table~\ref{tab:quadrupole}. The experimental value of the
isovector giant quadrupole resonance energy is just twice of the
isoscalar giant quadrupole resonance energy for $^{208}Pb$, and it
manifests the relation between the nuclear
isovector giant resonance and the corresponding isoscalar giant
resonance in Eq.~(\ref{twotime}) is correct.
Actually, the
experimental values for the nuclear giant quadrupole resonance of
other nuclei, and even for their monopole giant
resonance satisfy the relation in Eq.~(\ref{twotime}),
approximately.

In Table~\ref{tab:quadrupole}, Since the average
neutron density is larger than the saturation density of the nuclear
matter, the collective excitation energies of the nuclei $^{208}Pb$
and $^{90}Zr$ for $l=2$ are less than 10MeV. These values might
correspond to the low-lying excitation states in heavy
nuclei, which are in the range of $2-6$MeV for
$^{208}Pb$\cite{Ring-Schuck}.

\begin{table}
\tbl{The Fermi momenta and the $l=2$ collective excitation
energies of protons and neutrons for different nuclei.
The corresponding experimental values for the nuclear isoscalar
 giant quadrupole resonances $E^S_{exp}$ are taken from Ref.~\cite{Cao84}, and the
experimental values for the nuclear isovector giant quadrupole
resonances $E^V_{exp}$ from Refs.~\cite{Cao78,Cao}.}
{\begin{tabular}{@{}cccccc@{}}\toprule
$l=2$               & $^{208}Pb$& $^{90}Zr$ & $^{40}Ca$ & $^{16}O$\\
\colrule
$M^\ast_N/M_N$      &   0.742  &   0.742  & 0.69   & 0.69  \\
$k_F(p)~(fm^{-1})$  &  $1.26$  &   $1.31$ & $1.36$ &$1.36$  \\
 $k_F(n)~(fm^{-1})$ &  $1.45$  &   $1.41$ & $1.36$ &$1.36$  \\
 $E_0(p)~(MeV)$       &  $15.02$      &  $13.16$      &$18.54$      & $18.54$ \\
  $E_0(n)~(MeV)$      &  $5.84$       &   $8.27$      &$18.54$      &  $18.54$  \\
$E_0(p)+E_0(n)~(MeV)$ &  $20.86$      & $21.43$       &$37.08$      & $37.08$    \\
  $E^S_{exp}~(MeV)$   &  $10.9\pm0.1$ & $14.41\pm0.1$ &$17.8\pm0.3$ & $20.7$    \\
  $E^V_{exp}~(MeV)$   &  $22$         & $-$           &$32.5\pm1.5$ &    $-$  \\
\botrule
\end{tabular}
}
\label{tab:quadrupole}
\end{table}

The Fermi-liquid model in Ref.~\cite{Wen} is
extended to the 3-dimensional Fermion system with the spin
taken into account, and then by using the effective Lagrangian of the linear
$\sigma-\omega$ model, the Fermi-liquid function is obtained 
and the isoscalar and isovector giant resonances of some spherical
nuclei are studied. we find the
centroid energies of the isoscalar giant resonances just correspond
to the positive isolated energy levels of the nuclear collective
excitation with different values of $l$, respectively, while the
isovector giant resonances except $l=1$ correspond to the modes that
protons(neutrons) are in the creation state of the collective
excitation and neutrons(protons) are in the annihilation state of
the same $l$.
The low energy excitation of the nuclear giant quadrupole resonance,
might be described by the collective excitation of neutrons in the
nuclei.

It should be pointed out that the exchange interaction between nucleons plays an important role in the calculation
of collective excitations of nuclear systems by using a Fermi-liquid model.
Dirac sea of nucleons does not have any influence on the properties of giant resonances of nuclei.

This work is supported by the National Natural Science Foundation of
China under grant number 10775012.

\end{document}